\documentclass{elsart}
\usepackage{amssymb}
\journal{Physics Letters A}
\usepackage{graphicx}
\usepackage{dcolumn}
\usepackage{bm}
\usepackage{psfrag}
\usepackage{subfigure}

\newcommand{\be}{\begin{eqnarray}}
\newcommand{\ee}{\end{eqnarray}}
\newcommand{\bn}{\begin{eqnarray*}}
\newcommand{\en}{\end{eqnarray*}}

\newcommand{\nn}{\nonumber \\}
\newcommand{\nl}{\\}

\renewcommand{\vec}[1]{\mbox{\boldmath$#1$}}

\begin{document}

\begin{frontmatter}



\title{Neutrino wave packet propagation in gravitational fields}


\author[Regina]{Dinesh Singh},
\ead{singhd@uregina.ca}
\author[Regina]{Nader Mobed\corauthref{cor}},
\corauth[cor]{Corresponding author.} \ead{nader.mobed@uregina.ca}
\author[Regina,Salerno]{Giorgio Papini}
\ead{papini@uregina.ca}
\address[Regina]{Department of Physics, University of Regina, \\
Regina, Saskatchewan, S4S 0A2, Canada}
\address[Salerno]{International Institute for Advanced Scientific Studies, \\
89019 Vietri sul Mare (SA), Italy}

\begin{abstract}

We discuss the propagation of neutrino wave packets in a
Lense-Thirring space-time using a gravitational phase approach.
We show that the neutrino oscillation length is altered by
gravitational corrections and that neutrinos are subject to
helicity flip induced by stellar rotation.
For the case of a rapidly rotating neutron star, we show that absolute
neutrino masses can be derived, in principle, from rotational contributions
to the mass-induced energy shift, without recourse to mass
generation models presently discussed in the literature.

\end{abstract}

\begin{keyword}
neutrino wave packet \sep gravitational phase
\sep helicity flip \sep absolute neutrino mass

\PACS 04.90.+e \sep 14.60.Pq \sep 04.80.+z \sep 97.60.Bw
\end{keyword}
\end{frontmatter}


Recent experimental evidence from Superkamiokande \cite{Kam} and
SNO \cite{SNO} has significantly endorsed the claim \cite{gribov}
that neutrinos undergo flavour oscillations in vacuum due to the
difference of their rest masses. This discovery, however, makes no
pronouncements about the intrinsic physical properties of
neutrinos, particularly whether they exist as plane waves or wave
packets.
In addition, the actual calculation of the oscillation
length for solar neutrinos assumes a flat space-time background.
While this assumption seems reasonable for neutrino propagation in
our solar system, it may be unwarranted when the gravitational
source is a neutron star or a black hole.

The purpose of this work is to investigate the contribution of
gravitation to the flavour oscillations of a two-neutrino system
in a Lense-Thirring space-time background \cite{lense}, assuming a
wave packet description of the mass eigenstates in momentum space.
This approach \cite{giunti}, so far untested in gravitational
problems, offers an intuitively satisfying description of
neutrinos.
It relates directly the physical characteristics of
the gravitational source, such as its stellar temperature and density,
to the size of the wave packet.
It also lends itself well to sensing distortions in space-time because of the
common and intrinsically non-local nature of both curvature and
wave packets.
The choice of Lense-Thirring space-time is particularly relevant for our purpose,
since rotational effects may induce a helicity transition while the neutrino is in transit.

A particularly interesting feature of our approach is the introduction of the gravitational phase that leads to
a {\em direct} spin-gravity coupling interaction within the neutrino's wavefunction.
It was shown \cite{cai,singh,papini1,papini2} that the gravitational phase
\be
\Phi_{\rm G} & \equiv & \frac{1}{2} \int_{x_0}^{x} dz^\lambda
\gamma_{\alpha \lambda} (z) p^\alpha
 + \frac{1}{4} \int_{x_0}^{x} dz^\lambda \left[\gamma_{\beta \lambda,
\alpha} (z) - \gamma_{\alpha \lambda, \beta} (z) \right]
L^{\alpha \beta} (z),
\label{PhiG=}
\ee
enters the description of quantum particles in external
gravitational fields in a way that is essential and independent of
their spin. In (\ref{PhiG=}), $\gamma_{\mu \nu} = g_{\mu \nu} -
\eta_{\mu\nu}$ is the metric deviation, and $p^\mu$ and $L^{\alpha
\beta}$ are momentum and orbital angular momentum operators of the
free particle.
Application of (\ref{PhiG=}) to closed space-time
integration paths gives rise to a covariant Berry's phase
\cite{cai1} with consequences for particle interferometry
\cite{cai2}.
Equation (\ref{PhiG=}) also yields the particle
deflection predicted by general relativity in the geometrical
optics approximation \cite{lamb}. When applied to fermions, $\Phi_{\rm G}$
and the spin connection $\Gamma_{\mu}$ reproduce all
those gravitational-inertial effects that have been either
observed \cite{page,bonse} or predicted \cite{hehlni,mash},
and predicts, in particular, the non-conservation of helicity \cite{singh1} for strictly massless fermions.

To show how $\Phi_{\rm G}$  acts on a neutrino propagating in vacuum, we first start with the covariant Dirac equation
$\left[i \gamma^\mu (x) \left(\nabla_\mu + i \Gamma_\mu\right) - m/\hbar\right] \psi (x) = 0$,
where $m$ is the neutrino rest mass, and $\nabla_\mu$ is the usual covariant derivative. We use geometric units of $G = c = 1$
\cite{MTW}, so that all physical quantities can be described in units of length.
Then the corresponding Dirac Hamiltonian in Lense-Thirring space-time is
\be
H_0 & \approx & \left(1 - \frac{2M}{r}\right) \vec{\alpha} \cdot \vec{p}
+ m \left(1 - \frac{M}{r}\right) \beta + {i \hbar \, \frac{M}{2r^3}} \left(\vec{\alpha} \cdot \vec{r} \right)
\nn
& &{}+ \frac{4}{5} \, \frac{M\Omega R^2}{r^3} \, L^{\hat{z}} + \frac{1}{5} \, \frac{\hbar M \Omega R^2}{r^3}
\left[\frac{3z}{r^2} \left(\vec{\sigma} \cdot \vec{r}\right) - \sigma^{\hat{z}}\right],
\label{H}
\ee
where $r = \sqrt{x^2 + y^2 + z^2}$, $M$ and $R$ are the mass and radius of the gravitational source,
$\Omega$ is its angular velocity, and $L^{\hat{z}} = x p^{\hat{y}} - y p^{\hat{x}}$ is the orbital
angular momentum operator in the $z$-direction.
The gravitational phase can be introduced by means of the transformation
$\psi(x) \rightarrow \exp\left(i\Phi_{\rm G}/\hbar\right) \psi(x)$
and the new Hamiltonian takes the form
$H = H_0 
+ H_{\Phi_{\rm G}}$, where
\be
H_{\Phi_{\rm G}} & = & \vec{\alpha} \cdot \left(\vec{\nabla} \Phi_{\rm G} \right) +
\left(\vec{\nabla}_t \Phi_{\rm G} \right)
\label{H_phiG=}
\ee
is a first-order correction.
We then treat (\ref{H_phiG=}) as a perturbation for a two mass-species neutrino system.
While the approach taken is based on standard quantum mechanics, in order to apply (\ref{PhiG=})
it is necessary to assume that the average of all possible paths reduces to the integration path of $\Phi_{\rm G}$.
Because some terms of $H$ are non-diagonal with respect to spin, the wave packet approach should also make known whether the
neutrino mass eigenstates are subject to a helicity transition in the gravitational field of a rotating source.

Adopting the Dirac representation, the matrix element is
$\langle \psi(\vec{r}) |H_{\Phi_{\rm G}}| \psi(\vec{r}) \rangle$, and we assume for the wave packet description
\be
| \psi(\vec{r}) \rangle & = &
{1  \over (2\pi)^{3/2}} \int d^3 k \, \xi(\vec{k}) \,
e^{i \vec{k} \cdot \vec{r}} |U(\vec{k})\rangle_{\pm},
\label{psi=}
\ee
where $\xi(\vec{k}) = \xi(k^x)\xi(k^y)\xi(k^z)$ is the normalized Gaussian wavefunction of width
$\sigma_{\rm p}$ and centroid $\langle \vec{k} \rangle$, and
\be
\xi(k^j) & = & {1 \over ({\sqrt{2\pi} \, \sigma_{\rm p}})^{1/2}} \, \exp\left[- {(k^j - \langle \vec{k} \rangle^j)^2
\over 4 \sigma_{\rm p}^2} \right].
\label{Gaussian=}
\ee
The normalized four-spinor $|U(\vec{k})\rangle_{\pm}$ is
\be
|U(\vec{k})\rangle_{\pm} & = & \sqrt{E + m \over 2E}
\left(
\begin{array}{c}
1 \\
{\hbar \, \vec{\sigma} \cdot \vec{k} \over {E + m}}
\end{array} \right) \otimes | \pm \rangle, 
\nn
| + \rangle & = & \left(
\begin{array}{c}
1 \\
0
\end{array} \right), \qquad
| - \rangle \ = \ \left(
\begin{array}{c}
0 \\
1
\end{array} \right),
\label{Uspinor=}
\ee
where $E = \sqrt{p^2 + m^2}$ is the energy associated with the spinor.
By symmetrizing over the exchange between $\vec{k}$ and $\vec{k}'$, the explicit construction of the matrix element shows
contributions due to both a spin diagonal term and a helicity transition, in the form
\be
\lefteqn{\langle \psi(\vec{r}) |H_{\Phi_{\rm G}}| \psi(\vec{r}) \rangle \ = \ {\hbar^{3/2} \over (2\pi)^3 \, V} \int
d^3 r \ d^3 k \ d^3 k' \ \ \xi(\vec{k}) \, \xi(\vec{k}') \, \sqrt{E + m \over 2E} \sqrt{E' + m \over 2E'}}
\nn \nn \nn
&& \times \left\{\cos\left[(\vec{k} - \vec{k}') \cdot \vec{r}\right]
\left[\hbar \left(\vec{\nabla} \Phi_{\rm G} \right)_S \cdot \left({\vec{k} \over E + m}
+ {\vec{k}' \over E' + m}\right) \right. \right.
\nn
&& \left. + \left(\vec{\nabla}_t \Phi_{\rm G} \right)_S
\left(1 + {\hbar^2 \left(\vec{k} \cdot \vec{k}'\right) \over (E + m)(E' + m)} \right)\right]
\langle \pm | \pm \rangle
\nn \nn \nn
&& - \sin\left[(\vec{k} - \vec{k}') \cdot \vec{r}\right]
\left[\hbar \left(\vec{\nabla} \Phi_{\rm G} \right)_S \times \left({\vec{k} \over E + m}
- {\vec{k}' \over E' + m}\right) \right.
\nn
&& \left. \left. - {\hbar^2 \, \left(\vec{\nabla}_t \Phi_{\rm G} \right)_S
\left(\vec{k} \times \vec{k}'\right) \over (E + m)(E' + m)} \right] \cdot
\langle \mp | \vec{\sigma} | \pm \rangle \right\},
\label{H-matrix-element=}
\ee
where $\left(\vec{\nabla}_\mu \Phi_{\rm G} \right)_S = \left(\hbar/2\right)
\left[\left(\vec{\nabla}_\mu \Phi_{\rm G} \right) (\vec{k})
+ \left(\vec{\nabla}_\mu \Phi_{\rm G} \right) (\vec{k'})\right]$
is the symmetrized form of the gravitational phase gradients, and $V = \left({4 \pi/ 3}\right) \left(r^3 - R^3\right)$
is the volume of spatial integration from the star's surface.
It is clear from the last line of (\ref{H-matrix-element=}) that there would be no helicity transition contribution
if we considered strictly plane waves, since a non-zero transition amplitude requires different momentum components
within the wave packet to interact with the Pauli spin matrices to yield a non-zero result.
Choosing the spin quantization axis to be along the neutrinos' direction of propagation, the helicity transition element is
\be
\langle \mp | \vec{\sigma} | \pm \rangle & = &
\left[\cos \theta \, \cos \varphi \mp i \, \sin \varphi \right] \vec{\hat{x}} +
\left[\cos \theta \, \sin \varphi \pm i \, \cos \varphi \right] \vec{\hat{y}} -
\sin \theta \, \vec{\hat{z}},
\label{orientation=}
\ee
where the upper sign refers to the transition from negative to positive helicity.

The explicit calculation of the integrals of (\ref{H-matrix-element=}) in spherical coordinates is prohibitively complicated by the
coupling of the oscillatory functions to the amplitudes.
However, an approximate, but analytic expression
for (\ref{H-matrix-element=}) can be found by integrating
over both position and momentum space after performing a 2nd-order
Taylor series expansion with respect to $m$ and an 11th-order
Taylor expansion of the oscillatory functions.
It can then be shown that (\ref{H-matrix-element=}) has the form
\be
\langle \psi(\vec{r}) |H_{\Phi_{\rm G}}| \psi(\vec{r}) \rangle & = &
\hbar \langle k \rangle
\left\{
{M \over r} \left[C_0 (q,r) + C_1 (q,r) \, \bar{m} + C_2 (q,r) \, \bar{m}^2 \right] \right.
\nn
&  &{} + \left. {M \Omega R^2\over r^2} \sin \theta
\left[D_0 (q,r) + D_1 (q,r) \, \bar{m} + D_2 (q,r) \, \bar{m}^2 \right] \right\},
\nn
\label{amplitude1=}
\ee
where $\bar{m} = m/\langle p \rangle = m/(\hbar \langle k \rangle)$ and $q = \langle k \rangle/\sigma_{\rm p}$,
with $C_j(q,r)$ and $D_j(q,r)$ as dimensionless functions that can be tabulated.
The first line of (\ref{amplitude1=}) is the contribution due to the diagonal components of
$\langle \psi(\vec{r}) |H_{\Phi_{\rm G}}| \psi(\vec{r}) \rangle$,
while the second line is due to the off-diagonal components and
refers to the helicity flip contribution of the perturbation.
\begin{figure}
\psfrag{x1}[ll][][3][0]{$q$}
\psfrag{x2}[ll][][3][0]{$q$}
\psfrag{logC0}[bb][][4][90]{$\tiny \log_{10}\left[C_0(q,r)\right]$}
\psfrag{logC1}[bb][][4][90]{$\tiny \log_{10}\left[C_1(q,r)\right]$}
\psfrag{logC2}[bb][][4][90]{$\tiny \log_{10}\left[-C_2(q,r)\right]$}
\psfrag{logD0}[bb][][4][90]{$\tiny \log_{10}\left[D_0(q,r)\right]$}
\psfrag{logD1}[bb][][4][90]{$\tiny \log_{10}\left[-D_1(q,r)\right]$}
\psfrag{logD2}[bb][][4][90]{$\tiny \log_{10}\left[-D_2(q,r)\right]$}
\psfrag{logF0}[bb][][4][90]{$\tiny \log_{10}\left[F_0(q,r)\right]$}
\psfrag{logF1}[bb][][4][90]{$\tiny \log_{10}\left[-F_1(q,r)\right]$}
\psfrag{logF2}[bb][][4][90]{$\tiny \log_{10}\left[-F_2(q,r)\right]$}
\psfrag{2e-05}[ll][][2][0]{$0.2$}
\psfrag{4e-05}[ll][][2][0]{$0.4$}
\psfrag{6e-05}[ll][][2][0]{$0.6$}
\psfrag{8e-05}[ll][][2][0]{$0.8$}
\begin{minipage}[t]{0.3 \textwidth}
\centering
\subfigure[{\small $\quad C_0(q,r)$}]{
  \label{fig:C0}
\rotatebox{0}{\includegraphics[width = 4.5cm, height = 4.5cm, scale = 1]{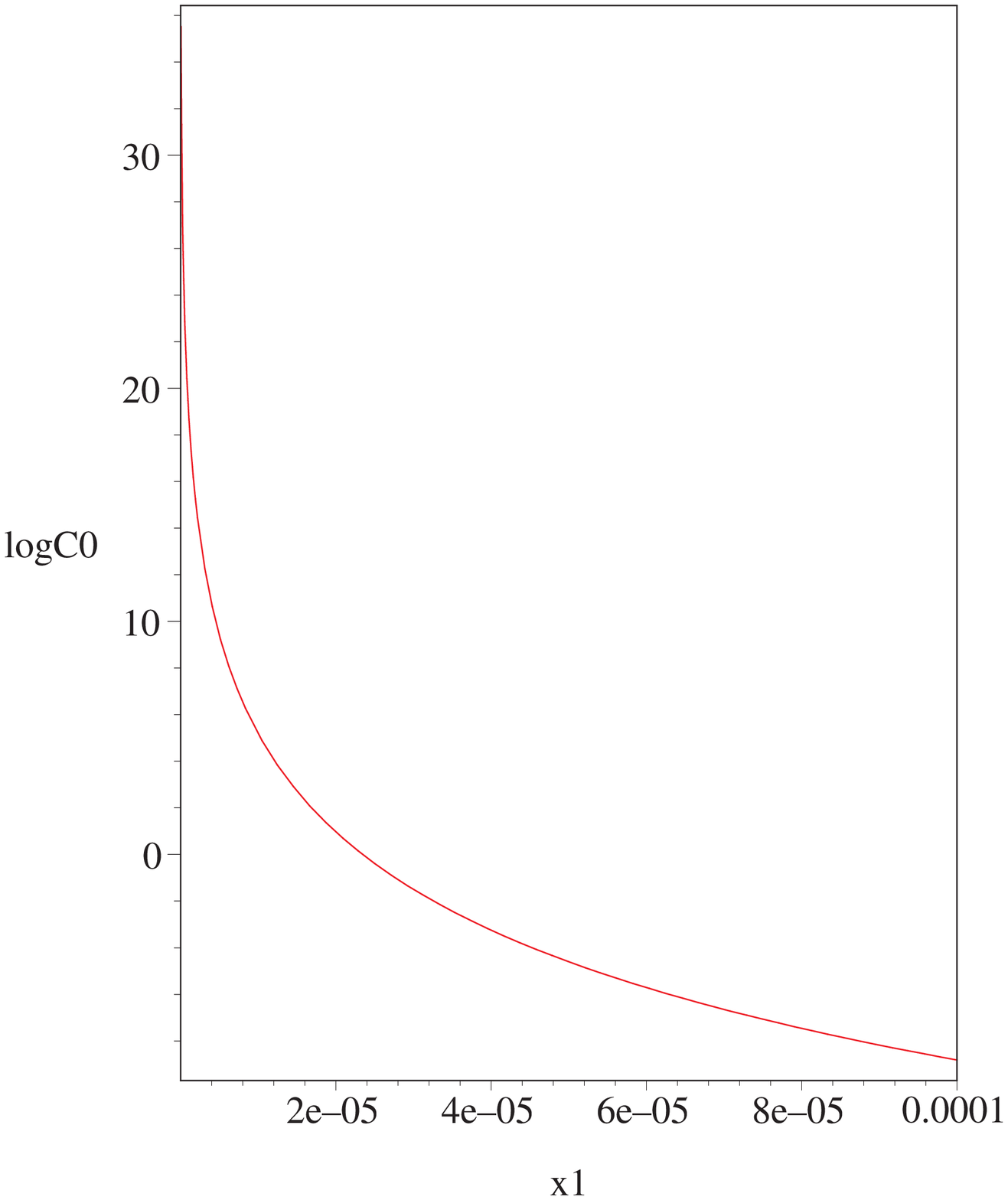}}}
\end{minipage}%
\hspace{0.5cm}
\begin{minipage}[t]{0.3 \textwidth}
\centering
\subfigure[{\small $\quad C_1(q,r)$}]{
  \label{fig:C1}
\rotatebox{0}{\includegraphics[width = 4.5cm, height = 4.5cm, scale = 1]{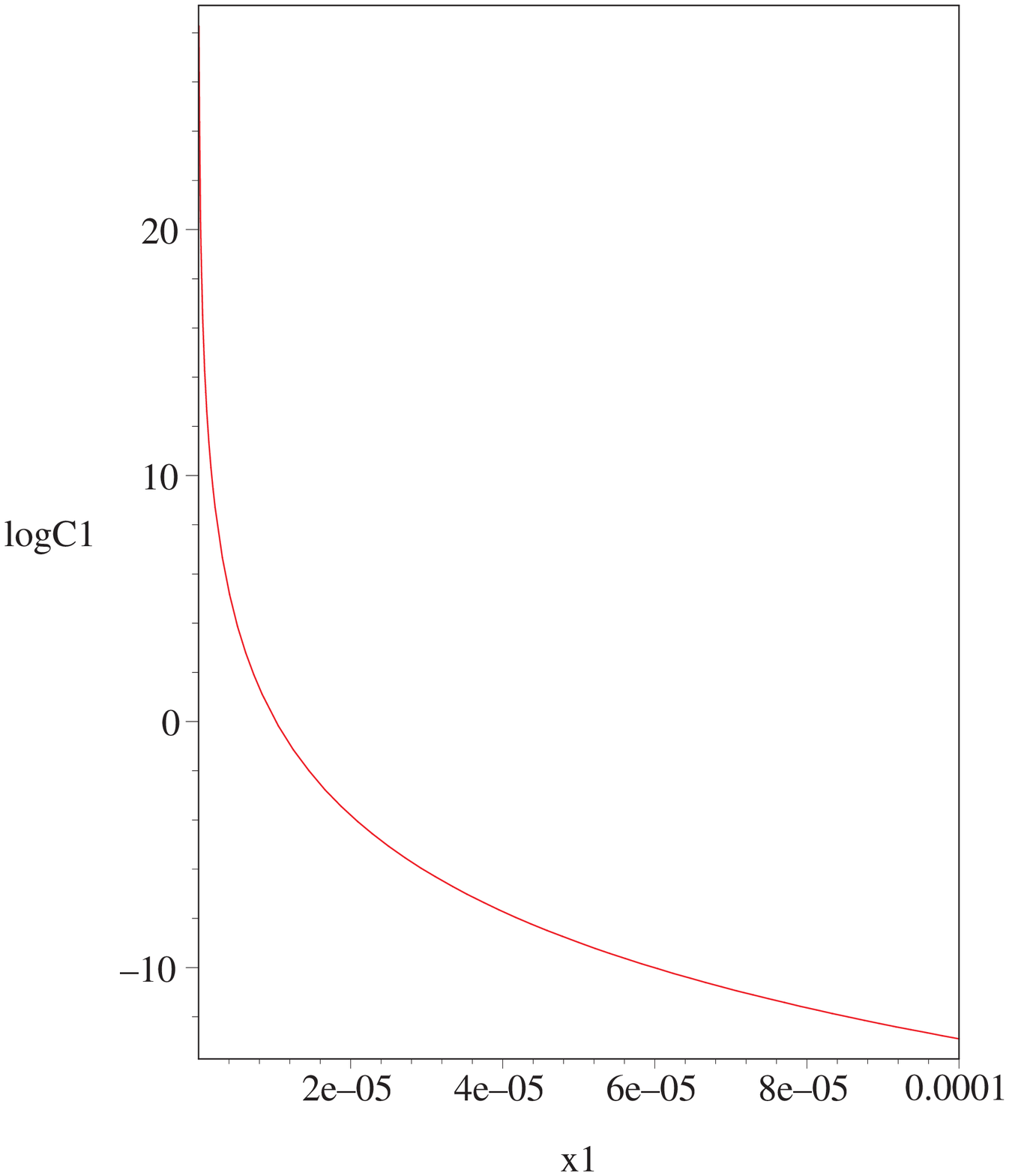}}}
\end{minipage}%
\hspace{0.5cm}
\begin{minipage}[t]{0.3 \textwidth}
\centering
\subfigure[{\small $\quad C_2(q,r)$}]{
  \label{fig:C2}
\rotatebox{0}{\includegraphics[width = 4.5cm, height = 4.5cm, scale = 1]{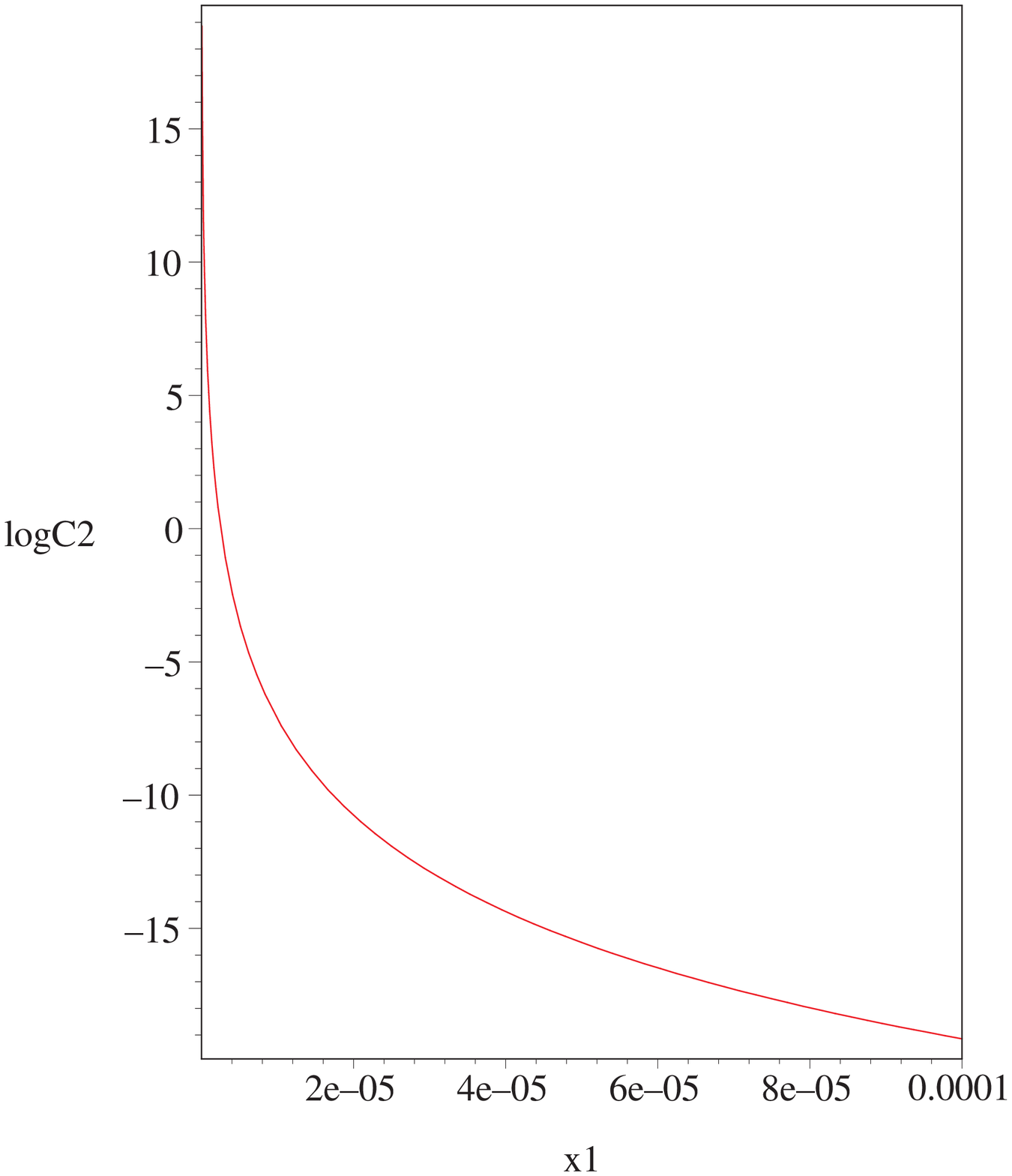}}}
\end{minipage}
\\
\begin{minipage}[t]{0.3 \textwidth}
\centering
\subfigure[{\small $\quad D_0(q,r)$}]{
  \label{fig:D0}
\rotatebox{0}{\includegraphics[width = 4.5cm, height = 4.5cm, scale = 1]{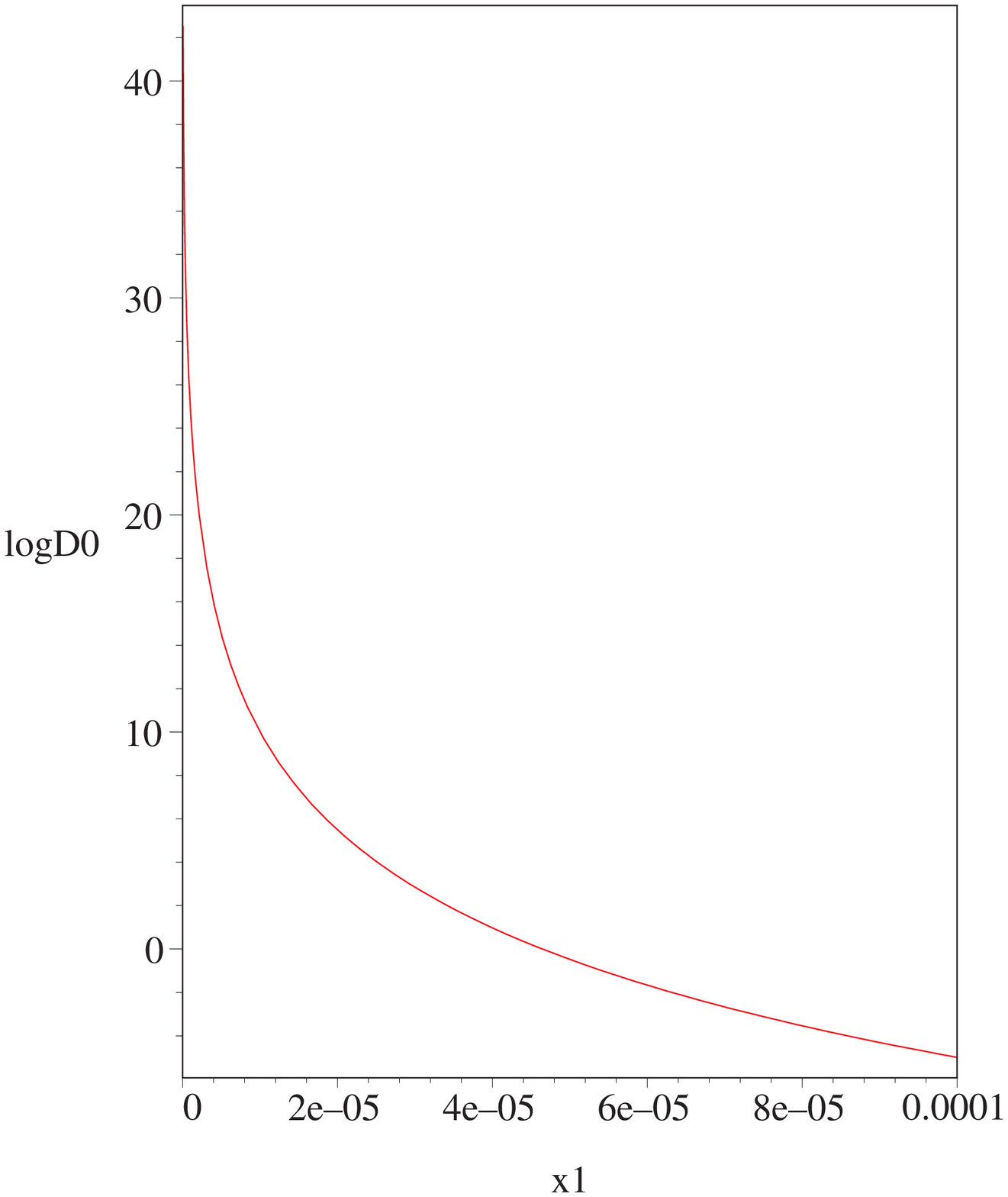}}}
\end{minipage}%
\hspace{0.5cm}
\begin{minipage}[t]{0.3 \textwidth}
\centering
\subfigure[{\small $\quad D_1(q,r)$}]{
  \label{fig:D1}
\rotatebox{0}{\includegraphics[width = 4.5cm, height = 4.5cm, scale = 1]{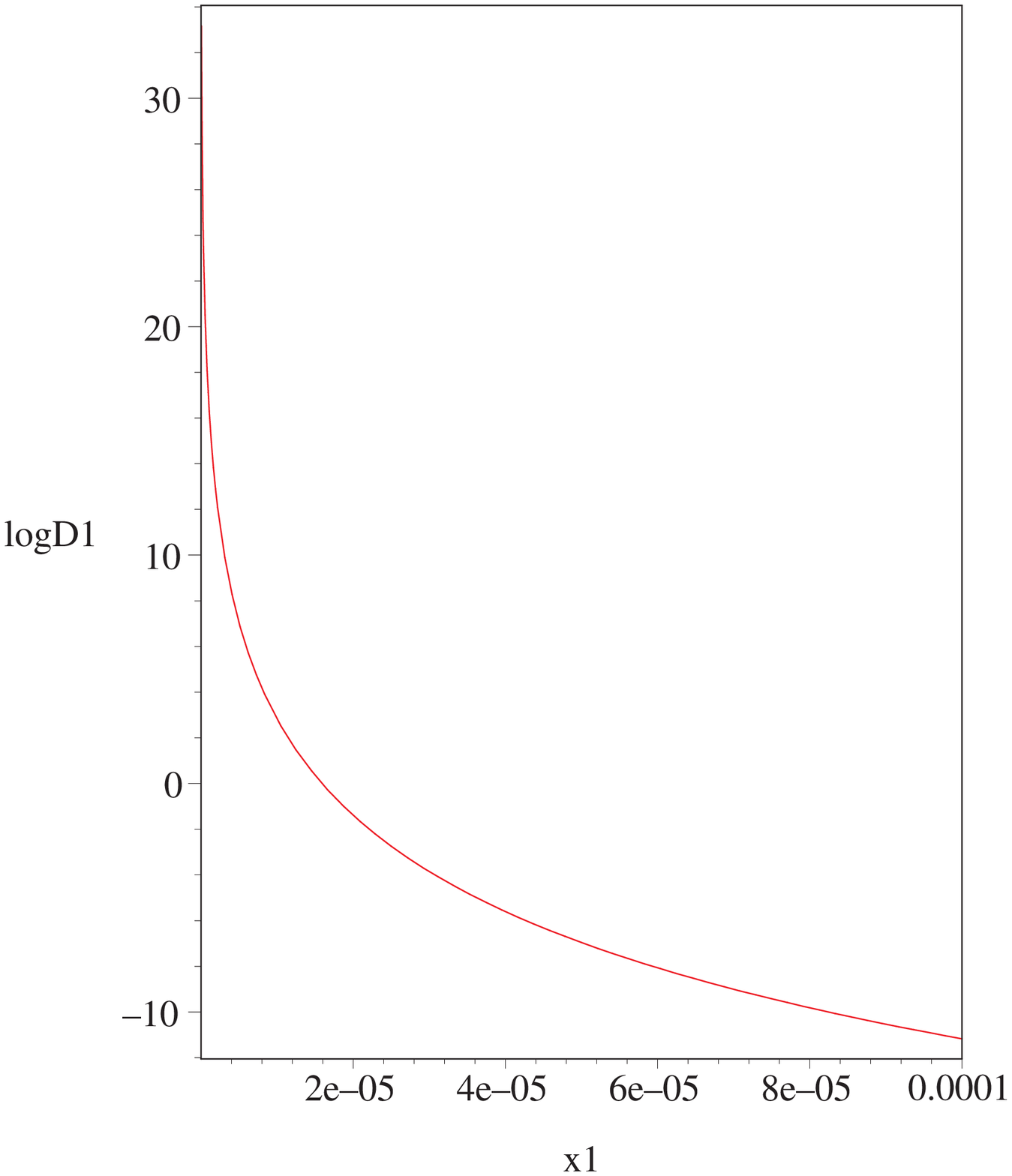}}}
\end{minipage}%
\hspace{0.5cm}
\begin{minipage}[t]{0.3 \textwidth}
\centering
\subfigure[{\small $\quad D_2(q,r)$}]{
  \label{fig:D2}
\rotatebox{0}{\includegraphics[width = 4.5cm, height = 4.5cm, scale = 1]{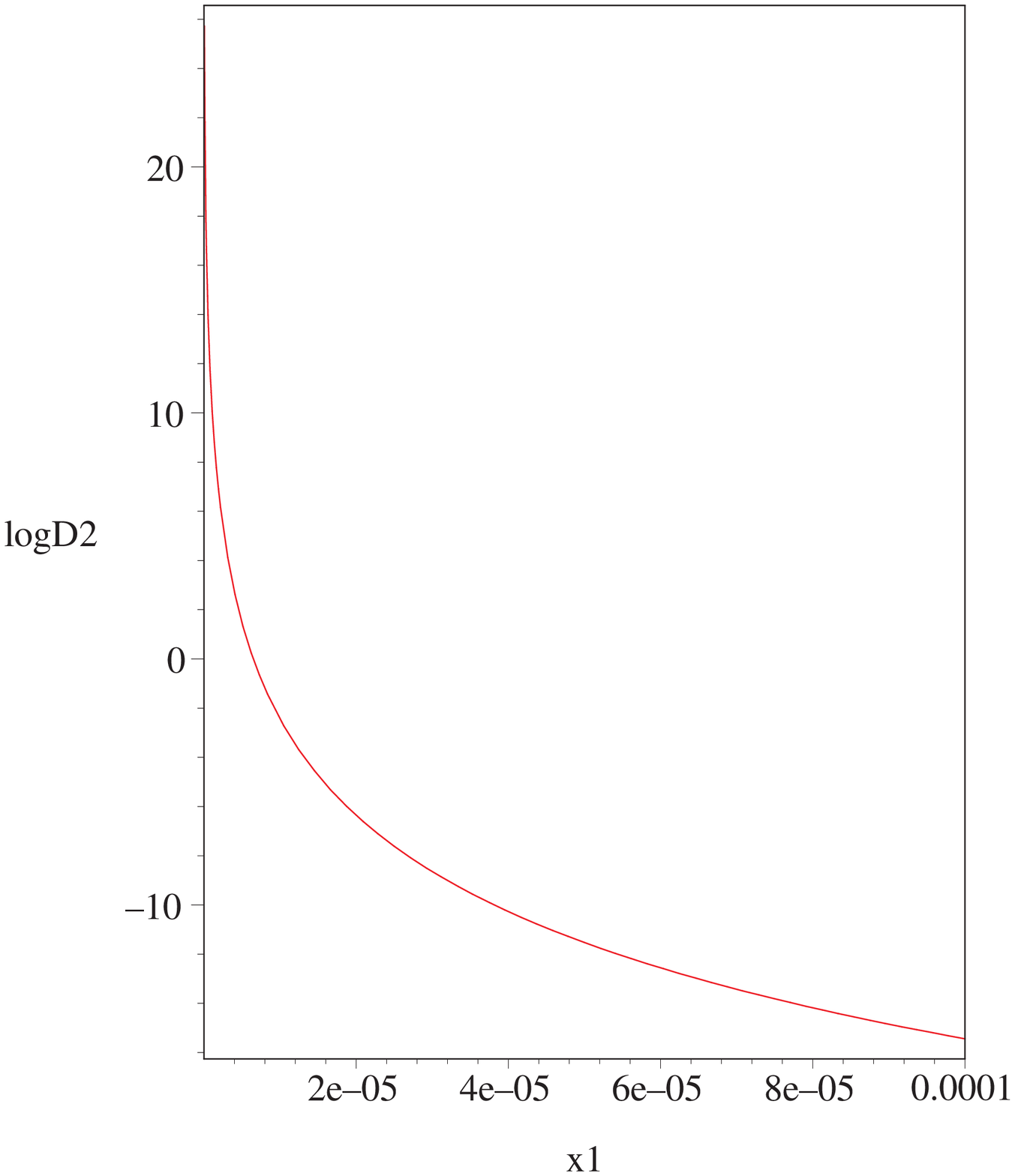}}}
\end{minipage}
\caption{\label{fig:coeff1} Functions which determine 
Eq. (\ref{amplitude1=}) for 10 MeV neutrinos,
given a $r = 10$ kpc neutron star source of $M = 1.5 M_{\odot}$ and $R = 10$ km.}
\end{figure}

For this paper, the gravitational source under consideration is a rapidly rotating
$1.5 M_{\odot}$ neutron star with $R = 10$ km and $\Omega = 1$ kHz, at a distance of $r = 10$ kpc.
Figure~\ref{fig:coeff1} contains a list of plots for the functions described in the matrix element (\ref{amplitude1=}).
For this choice of parameters, it is clear from a comparison of Figures~\ref{fig:C0}-\ref{fig:C2}
and Figures~\ref{fig:D0}-\ref{fig:D2} that
$|C_2(q,r)| \ll |C_1(q,r)| \ll |C_0(q,r)|$ and $|D_2(q,r)| \ll |D_1(q,r)| \ll |D_0(q,r)|$ for all choices of $q$.
This suggests the trend towards convergence of the series due to expansion with respect to $m$.
To justify the truncation of the Taylor expansion of (\ref{H-matrix-element=}),
we note from analyzing a one-dimensional analogue of the problem that the Gaussian functions
which are present in the integrand have the effect of damping out the contribution of the
higher-order expansion terms beyond the 11th order.
A plot of this series expansion matches precisely with that due to the corresponding product function
of the exact sinusoidal function with a one-dimensional Gaussian, except in the tail
regions of the Gaussian envelope.
In those isolated regions of parameter space, we estimate an error of 10-15\% in the amplitude
and are confident that this degree of error has no significant bearing on our results.

We stress some interesting features of (\ref{amplitude1=}).
The most obvious one is the presence of terms {\em linear} in $m$, due to the Taylor expansion of
$\langle \psi(\vec{r}) |H_{\Phi_{\rm G}}| \psi(\vec{r}) \rangle$ for both the spin diagonal and off-diagonal terms.
This fact has interesting implications, as shown below, in the calculation of the energy shifts.
Given (\ref{orientation=}), we know that the helicity transition contribution to (\ref{amplitude1=}) is due entirely to the
$z$-component of the transition amplitude, since the $x$- and $y$-components average out to zero over all spatial angles.
Furthermore, the non-zero contribution is due entirely to the rotation of the source, which induces the helicity transition.
Because of the $\sin \theta$ term, propagation of a neutrino along the $\pm z$-axis suggests that its initially prepared helicity
state remains fixed throughout its motion.
This is a sensible result that is consistent with our present understanding of spin-rotation coupling, since the rotational term in the
Lense-Thirring metric tends to act like a uniform magnetic field that orients the particle spin either parallel or antiparallel to
the field strength. This result also agrees with the fact that, barring quantum fluctuations about the classical integration path
of (\ref{PhiG=}), $\vec{\nabla}_t \Phi_{\rm G} = 0$ in the Lense-Thirring field, which implies helicity conservation \cite{singh1}.

\begin{figure}
\psfrag{x1}[ll][][3][0]{$q$}
\psfrag{x2}[ll][][3][0]{$q$}
\psfrag{logC0}[bb][][4][90]{$\tiny \log_{10}\left[C_0(q,r)\right]$}
\psfrag{logC1}[bb][][4][90]{$\tiny \log_{10}\left[C_1(q,r)\right]$}
\psfrag{logC2}[bb][][4][90]{$\tiny \log_{10}\left[-C_2(q,r)\right]$}
\psfrag{logD0}[bb][][4][90]{$\tiny \log_{10}\left[D_0(q,r)\right]$}
\psfrag{logD1}[bb][][4][90]{$\tiny \log_{10}\left[-D_1(q,r)\right]$}
\psfrag{logD2}[bb][][4][90]{$\tiny \log_{10}\left[-D_2(q,r)\right]$}
\psfrag{logF0}[bb][][4][90]{$\tiny \log_{10}\left[F_0(q,r)\right]$}
\psfrag{logF1}[bb][][4][90]{$\tiny \log_{10}\left[-F_1(q,r)\right]$}
\psfrag{logF2}[bb][][4][90]{$\tiny \log_{10}\left[-F_2(q,r)\right]$}
\psfrag{2e-05}[ll][][2][0]{$0.2$}
\psfrag{4e-05}[ll][][2][0]{$0.4$}
\psfrag{6e-05}[ll][][2][0]{$0.6$}
\psfrag{8e-05}[ll][][2][0]{$0.8$}
\begin{minipage}[t]{0.3 \textwidth}
\centering
\subfigure[{\small $\quad F_1(q,r)$}]{
  \label{fig:F1}
\rotatebox{0}{\includegraphics[width = 6cm, height = 6cm, scale = 1]{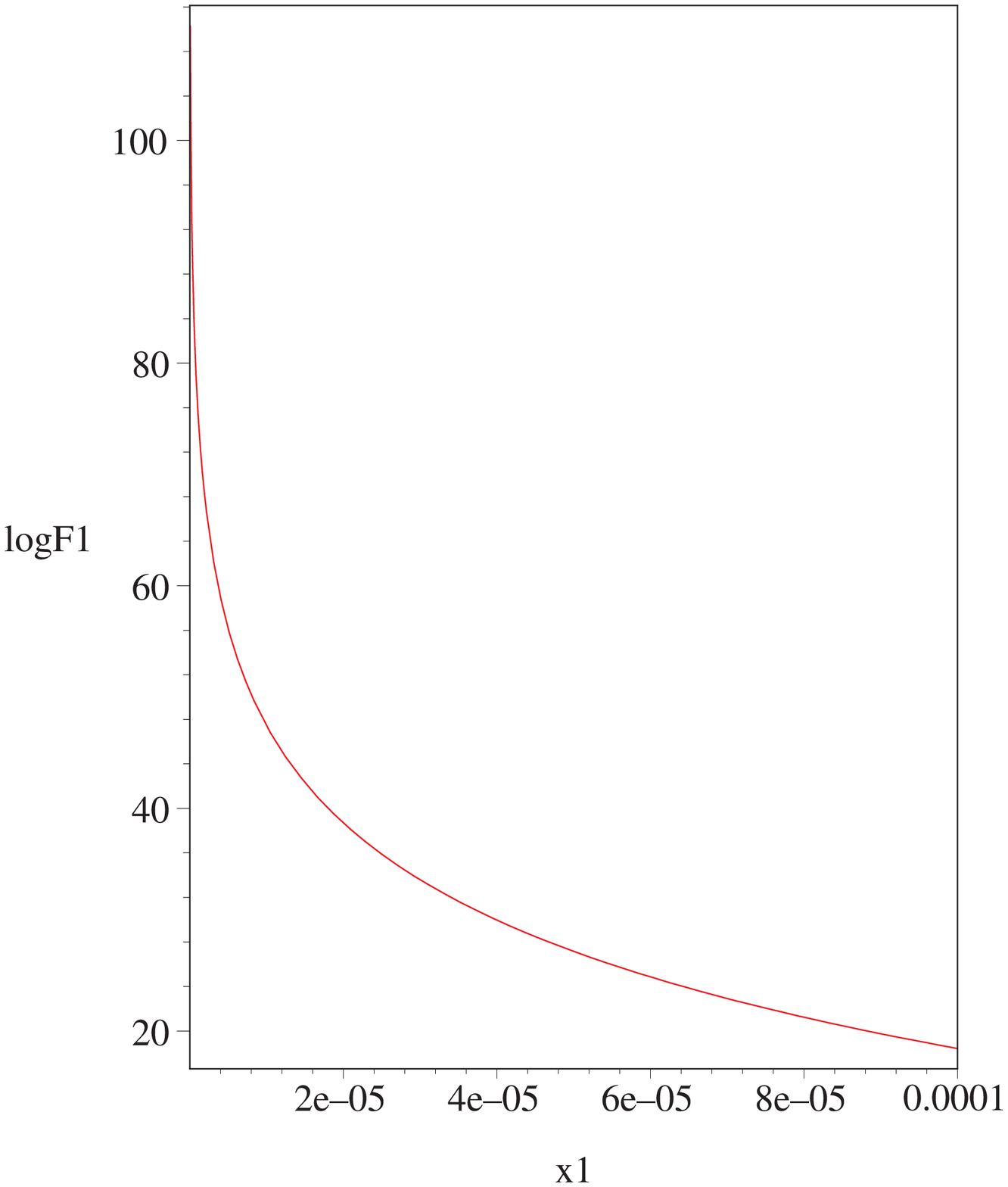}}}
\end{minipage}%
\hspace{3cm}
\begin{minipage}[t]{0.3 \textwidth}
\centering
\subfigure[{\small $\quad F_2(q,r)$}]{
  \label{fig:F2}
\rotatebox{0}{\includegraphics[width = 6cm, height = 6cm, scale = 1]{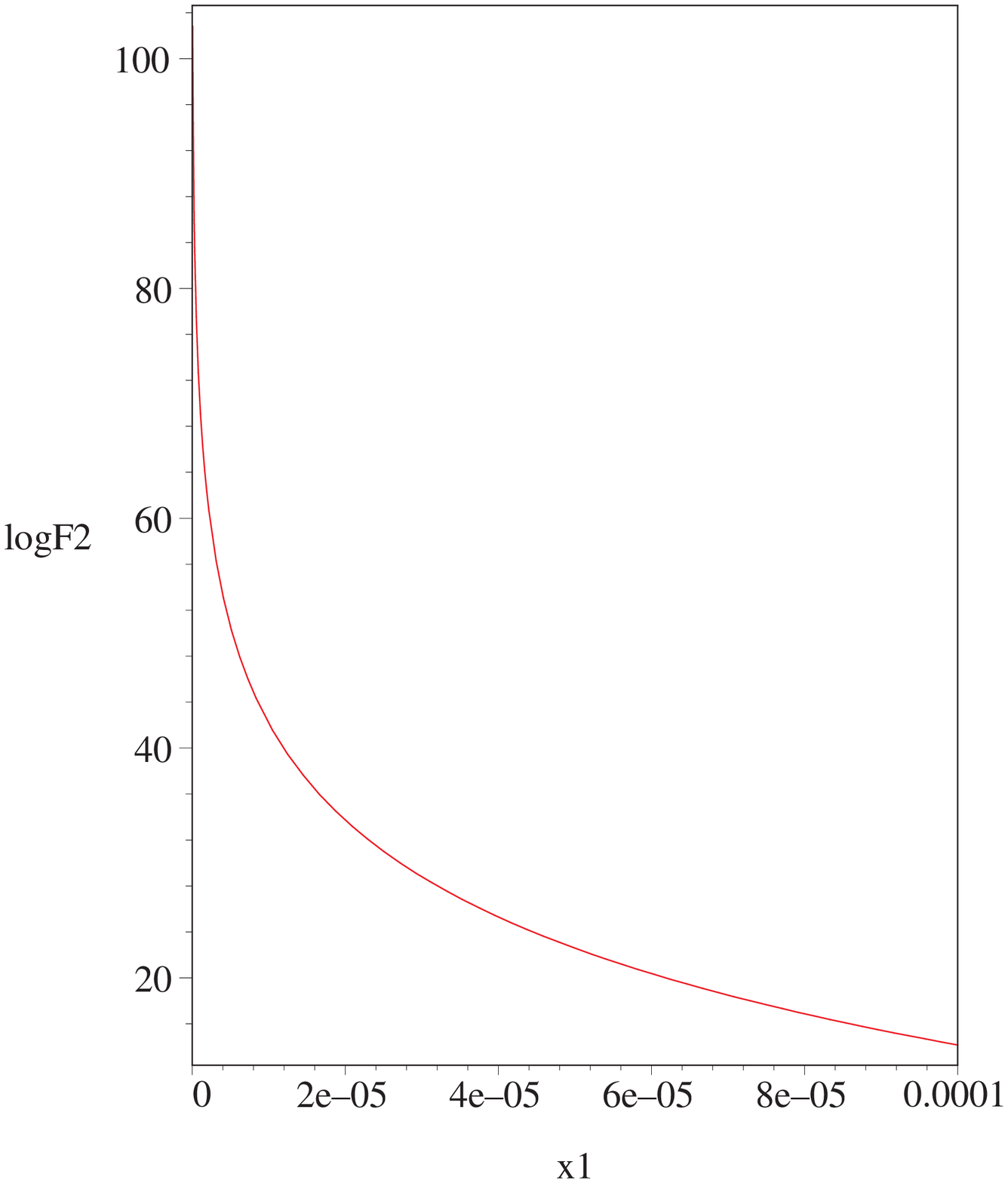}}}
\end{minipage}
\caption{\label{fig:coeff2} Corresponding functions for 
Eq. (\ref{Energy-shift-polarized=}), assuming the same source and neutrino energy conditions.}
\end{figure}

To determine the mass-induced energy shift for the neutrino oscillation length, we note that the off-diagonal elements of
(\ref{amplitude1=}) contribute to a second-order effect in time-independent perturbation theory, where the unperturbed
energy eigenvalues $E_0^{\pm}$ come from $H_0 \, \psi(x) = E_0^{\pm} \, \psi(x)$ for
$E_0^{\pm} \approx \sqrt{\langle p \rangle^2 + m^2} - {2M \over r} \, \langle p \rangle
+ {4 \over 5} \,  {M \Omega R^2\over r^3} \left(L^{\hat{z}} \pm {\hbar \over 2}\right)$,
and
\be
E_0^+ - E_0^- = {4 \over 5} \, {\hbar M \Omega R^2 \over r^3}.
\label{deltaE0=}
\ee
By virtue of (\ref{deltaE0=}), we can calculate the energy shift for a given neutrino, leading to a new energy of
$E_{\bar{m}}^{\pm} \approx E_0^{\pm} + \left(\Delta E\right)_{\bar{m}}^{\pm}$, where
\be
\left(\Delta E\right)_{\bar{m}}^{\pm} & = &
\langle \pm |H_{\Phi_{\rm G}}| \pm \rangle \pm
{\left| \langle - |H_{\Phi_{\rm G}}| + \rangle \right|^2 \over
E_0^+ - E_0^-}.
\label{2nd-perturbation=}
\ee
This leads \cite{fukugita,kim} to the final expression for the
neutrino oscillation length $L_{\rm osc.} = 2
\pi/\left(E_{\bar{m}_2}^{\pm} - E_{\bar{m}_1}^{\pm}\right)$, where
\be
E_{\bar{m}_2}^{\pm} - E_{\bar{m}_1}^{\pm} & = &
\hbar \langle k \rangle \left\{ {1 \over 2} \left(\bar{m}_2^2 -  \bar{m}_1^2\right) \right.
\nn
& &{} + \left[{M \over r} \, C_1(q,r) \pm
{M \Omega R^2\over r^2} \, \sin^2 \theta \, F_1(q,r) \right]
\left(\bar{m}_2 -  \bar{m}_1\right)
\nn
&  &{} + \left. \left[{M \over r} \, C_2(q,r) \pm
{M \Omega R^2\over r^2} \, \sin^2 \theta \, F_2(q,r) \right]
\left(\bar{m}_2^2 -  \bar{m}_1^2\right) \right\}
\label{Energy-shift-polarized=}
\ee
and
%
\be
F_0(q,r) & = & {5 \over 4} \, r \langle k \rangle \, D_0^2(q,r)
\label{F0}
\nl
F_1(q,r) & = & {5 \over 2} \, r \langle k \rangle \, D_0(q,r) \, D_1(q,r)
\label{F1}
\nl
F_2(q,r) & = & {5 \over 4} \, r \langle k \rangle \left[D_1^2(q,r) + 2 D_0(q,r) \, D_2(q,r)\right].
\label{F2}
\ee
%
The plots of (\ref{F1}) and (\ref{F2}) are listed in Figure~\ref{fig:coeff2}.


It is clear from Figures~\ref{fig:coeff1} and \ref{fig:coeff2}
that the functions become extremely large for $q \rightarrow 0$,
corresponding to very large momentum spread in the wave
packet for a given neutrino energy, which then rapidly decay to
zero for large $q$.
In particular, the plots show that the
helicity transition terms in (\ref{Energy-shift-polarized=}) will
dominate as $q \rightarrow 0$, which is consistent with
(\ref{H-matrix-element=}), since {\it a large momentum spread in
the wave packet is required to sense the differential rotational
effects within a localized region of space-time near the star's
surface}.

For solar neutrinos and those due to supernovae \cite{kim}, the
expected wave packet widths in momentum space are $\hbar
\sigma_{\rm p} \approx 10^{-5}$ MeV and $2 \times 10^{-2}$ MeV,
respectively.
The corresponding values for $q \approx 10^6$ and $q
\approx 50$ indicate that gravitational corrections have
negligible effect on the neutrino oscillation lengths for these
scenarios.
However, the situation is quite different when applied to
rapidly rotating neutron stars.
For the neutron star parameters we consider, ${M/r} \approx 7.175 \times
10^{-18}$ and ${M \Omega R^2/r^2} \approx 7.751 \times 10^{-36}$.
In order to predict a 1\% correction to the value of
$\Delta m_{21}^2 = m_2^2 - m_1^2$ as determined by solar neutrino
experiments, we require $|F_2(q,r)| \approx 1.3667 \times
10^{33}$, which suggests a choice of $q \approx 2.1 \times
10^{-5}$ from Figure~\ref{fig:F2}, and implies a wave packet width
of $\hbar \sigma_{\rm p} \approx 4.762 \times 10^5$ MeV.
This prediction corresponds very well to the calculated width of $\hbar
\sigma_{\rm p} \approx 3.260 \times 10^5$ MeV for neutrinos
emitted from neutron stars, as determined by a mean-free-path
calculation \cite{kim}, assuming a stellar temperature of $3 \times
10^6$ K \cite{shapiro}, and an effective stellar density of
$10^{11}$ g/cm$^3$ averaged over the neutron star's expected core
and surface densities.
Our results therefore show that helicity
flip likely plays a role in the case of neutrinos propagating in
the field of rotating neutron stars.

From Figure~\ref{fig:F1}, it also follows that $|F_1(q,r)| \approx
1.2150 \times 10^{38}$ for the same choice of $q$ as when applied to
Figure~\ref{fig:F2}, and so the contribution in
(\ref{Energy-shift-polarized=}) due to linear mass difference
$\Delta m_{21} = m_2 - m_1$ is {\em not} negligible. This result
suggests the possibility of performing, in principle, a parameter
fit of (\ref{Energy-shift-polarized=}) for suitable choices of
$q$, $\Delta m_{21}^2$, and $\Delta m_{21}$ so that we can infer
the absolute neutrino masses entirely from observations, and
without any reference to the mass generation mechanisms presently
under consideration in the literature \cite{fukugita}.
In practice, however, such an undertaking would require an
enormously large neutrino flux and large counting rates to
obtain statistically significant measurements.
To our knowledge, there are no experimental facilities available that
can meet these requirements.
Nonetheless, the theoretical possibility of determining absolute
neutrino masses by this technique makes for an interesting
consideration in the development of future neutrino observatories.

\section{Acknowledgements}

This research was supported, in part, by the Natural Sciences and Engineering Research Council of Canada (NSERC).

\end{document}